\documentstyle[12pt]{article}
\catcode`\@=11
\long\def\@makefntext#1{
\protect\noindent \hbox to 3.2pt {\hskip-.9pt
$^{{\ninerm\@thefnmark}}$\hfil}#1\hfill}		%CAN BE USED

\def\@makefnmark{\hbox to 0pt{$^{\@thefnmark}$\hss}}  %ORIGINAL

\def\ps@myheadings{\let\@mkboth\@gobbletwo
\def\@oddhead{\hbox{}
\rightmark\hfil\ninerm\thepage}
\def\@oddfoot{}\def\@evenhead{\ninerm\thepage\hfil
\leftmark\hbox{}}\def\@evenfoot{}
\def\sectionmark##1{}\def\subsectionmark##1{}}

\renewcommand{\thefootnote}{\fnsymbol{footnote}}

\newcounter{sectionc}\newcounter{subsectionc}\newcounter{subsubsectionc}
\renewcommand{\section}[1] {\vspace*{0.6cm}\addtocounter{sectionc}{1}
\setcounter{subsectionc}{0}\setcounter{subsubsectionc}{0}\noindent
	{\normalsize\bf\thesectionc. #1}\par\vspace*{0.4cm}}
\renewcommand{\subsection}[1] {\vspace*{0.6cm}\addtocounter{subsectionc}{1}
	\setcounter{subsubsectionc}{0}\noindent
	{\normalsize\it\thesectionc.\thesubsectionc. #1}\par\vspace*{0.4cm}}
\renewcommand{\subsubsection}[1]
{\vspace*{0.6cm}\addtocounter{subsubsectionc}{1}
	\noindent {\normalsize\rm\thesectionc.\thesubsectionc.\thesubsubsectionc.
	#1}\par\vspace*{0.4cm}}

\newcounter{appendixc}
\newcounter{subappendixc}[appendixc]
\newcounter{subsubappendixc}[subappendixc]

\renewcommand{\appendix}[1] {\vspace*{0.6cm}
        \refstepcounter{appendixc}
        \setcounter{figure}{0}
        \setcounter{table}{0}
        \setcounter{equation}{0}
        \renewcommand{\thefigure}{\Alph{appendixc}.\arabic{figure}}
        \renewcommand{\thetable}{\Alph{appendixc}.\arabic{table}}
        \renewcommand{\theappendixc}{\Alph{appendixc}}
        \renewcommand{\theequation}{\Alph{appendixc}.\arabic{equation}}
        \noindent{\bf Appendix \theappendixc #1}\par\vspace*{0.4cm}}

\def\abstracts#1{{

\centering{\begin{minipage}{12.2truecm}\footnotesize\baselineskip=12pt\noindent
	\centerline{\footnotesize ABSTRACT}\vspace*{0.3cm}
	\parindent=0pt #1
	\end{minipage}}\par}}

\renewenvironment{thebibliography}[1]
	{\begin{list}{\arabic{enumi}.}
	{\usecounter{enumi}\setlength{\parsep}{0pt}
\setlength{\leftmargin 1.25cm}{\rightmargin 0pt}
	 \setlength{\itemsep}{0pt} \settowidth
	{\labelwidth}{#1.}\sloppy}}{\end{list}}

\newcounter{itemlistc}
\newcounter{romanlistc}
\newcounter{alphlistc}
\newcounter{arabiclistc}

\newcommand{\fcaption}[1]{
        \refstepcounter{figure}
        \setbox\@tempboxa = \hbox{\footnotesize Fig.~\thefigure. #1}
        \ifdim \wd\@tempboxa > 6in
           {\begin{center}
        \parbox{6in}{\footnotesize\baselineskip=12pt Fig.~\thefigure. #1}
            \end{center}}
        \else
             {\begin{center}
             {\footnotesize Fig.~\thefigure. #1}
              \end{center}}
        \fi}

\newcommand{\tcaption}[1]{
        \refstepcounter{table}
        \setbox\@tempboxa = \hbox{\footnotesize Table~\thetable. #1}
        \ifdim \wd\@tempboxa > 6in
           {\begin{center}
        \parbox{6in}{\footnotesize\baselineskip=12pt Table~\thetable. #1}
            \end{center}}
        \else
             {\begin{center}
             {\footnotesize Table~\thetable. #1}
              \end{center}}
        \fi}

\def\@citex[#1]#2{\if@filesw\immediate\write\@auxout
	{\string\citation{#2}}\fi
\def\@citea{}\@cite{\@for\@citeb:=#2\do
	{\@citea\def\@citea{,}\@ifundefined
	{b@\@citeb}{{\bf ?}\@warning
	{Citation `\@citeb' on page \thepage \space undefined}}
	{\csname b@\@citeb\endcsname}}}{#1}}

\newif\if@cghi
\def\cite{\@cghitrue\@ifnextchar [{\@tempswatrue
	\@citex}{\@tempswafalse\@citex[]}}
\def\citelow{\@cghifalse\@ifnextchar [{\@tempswatrue
	\@citex}{\@tempswafalse\@citex[]}}
\def\@cite#1#2{{$\null^{#1}$\if@tempswa\typeout
	{IJCGA warning: optional citation argument
	ignored: `#2'} \fi}}

 1
 1
 1

\font\ninerm=cmr9

\newcommand{\apriori}{{\em a priori\/}}
\newcommand{\etal}{{\em et al.}}
\newcommand{\eg}{{\em e.g.}}

\newcommand{\beq} {\begin{equation}}
\newcommand{\eeq} {\end{equation}}
\newcommand{\bea} {\begin{eqnarray}}
\newcommand{\eea} {\end{eqnarray}}

\newcommand{\nc}{\newcommand}
\newcommand{\rnc}{\renewcommand}

\nc{\rome}{{\rm Roma}}
\nc{\ie}{{\em i.e.}}
\nc{\calF}{{\cal F}}
\nc{\calS}{{\cal S}}
\nc{\kapbar}{\bar{\kappa}}
\nc{\uidot}{\dot u}
\nc{\uiidot}{\ddot u}
\nc{\uiiidot}{\stackrel{\ldots}{u}}
\nc{\psii}{\psi^{(1)}}
\nc{\psibari}{\bar\psi^{(1)}}

\rnc{\topfraction}{1.0}
\rnc{\bottomfraction}{1.0}
\rnc{\textfraction}{0.0}

\nc{\qq}{\P}
\nc{\qqc}{[\P]}

\nc{\rng}{\rangle}
\nc{\lng}{\langle}
\nc{\rcite}{ref.\ \cite}
\nc{\ba}{\begin{array}}
\nc{\ea}{\end{array}}
\nc{\lb}{\left(}
\nc{\rb}{\right)}
\nc{\qrt}{\frac{1}{4}}

\nc{\al}{\alpha}
\nc{\bt}{\beta}
\nc{\gm}{\gamma}
\nc{\dl}{\delta}
\nc{\ep}{\epsilon}
\nc{\varep}{\varepsilon}
\nc{\zt}{\zeta}
\nc{\et}{\eta}
\nc{\th}{\theta}
\nc{\kp}{\kappa}
\nc{\lm}{\lambda}
\nc{\rh}{\rho}
\nc{\sg}{\sigma}
\nc{\ta}{\tau}
\nc{\ph}{\phi}
\nc{\vr}{\varphi}
\nc{\ch}{\chi}
\nc{\ps}{\psi}
\nc{\om}{\omega}
\nc{\noi}{\noindent}
\nc{\rr}[1]{$^{#1}$}
\nc{\rf}[1]{(\ref{#1})}
\nc{\rfs}[2]{(\ref{#1},\ref{#2})}
\nc{\smgr}{\stackrel{\textstyle <}{>}}
\nc{\grsm}{\stackrel{\textstyle >}{<}}
\nc{\aleq}{\ \mbox{}_{\textstyle \sim}^{\textstyle < }\ }
\nc{\ageq}{\ \mbox{}_{\textstyle \sim}^{\textstyle > }\ }
\nc{\ra}{\rightarrow}
\nc{\lra}{\leftrightarrow}
\nc{\be}{\begin{equation}}
\nc{\ee}{\end{equation}}
\nc{\eqrf}{eq.\ \rf}
\nc{\erf}{{\rm erf}}

\begin{document}

\input epsf

\hfill DFTUZ/95/17\\
\hspace*{\fill} hep-th/9505019\\
\hspace*{\fill} April 1995\\[1cm]

\centerline{\normalsize\bf SOME VIEWS ON MONOPOLES AND
CONFINEMENT\protect\footnote{Invited talk at the International RCNP
Workshop on COLOR CONFINEMENT AND HADRONS --- CONFINEMENT 95,
March 22--24, 1995, RCNP Osaka, Japan.}}
\baselineskip=16pt
\vspace*{0.6cm}
\centerline{\footnotesize A.J. van der Sijs}
\vspace*{0.4cm}
\baselineskip=13pt
\centerline{\footnotesize\it Department of Theoretical Physics,
    University of Zaragoza}
\baselineskip=12pt
\centerline{\footnotesize\it
    Facultad de Ciencias, 50009\ \ Zaragoza, Spain}
\centerline{\footnotesize E-mail: arjan@sol.unizar.es}
\vspace*{0.9cm}
\abstracts{Aspects of the monopole condensation picture of
confinement are discussed.
First, the nature of the monopole singularities in the abelian
projection approach is analysed.
Their apparent gauge dependence
is shown to have
a natural interpretation in terms of 't~Hooft-Polyakov-like monopoles
in euclidean SU(2) gauge theory.
Next, the results and predictions of a realization of confinement through
condensation of such monopoles are summarized and compared with
numerical data.}

\normalsize\baselineskip=14pt
\setcounter{footnote}{0}
\renewcommand{\thefootnote}{\alph{footnote}}
\section{Introduction}

Much of the effort in the area of confinement through monopole condensation
(the dual superconductor hypothesis\cite{Ho,Ma}) in recent years has been
motivated by the abelian projection approach.\cite{Ho81}
In this approach, the presence of magnetic monopoles in non-abelian
gauge theories is made visible by means of a partial gauge fixing to an
``abelian gauge'', and condensation of such monopoles might explain
quark confinement.
The nature of these monopoles has remained somewhat obscure,
however, due to the gauge field singularity characterizing them
and to the apparent dependence on the choice of abelian gauge.

One aim of this talk is to emphasize, following the original work,\cite{Ho81}
that the monopoles in this approach are physically relevant dynamical
variables.  I shall argue that they are regular extended objects with
semiclassical properties such as a size and a mass or action density
associated to them, which do not depend on the chosen abelian
gauge, even though the gauge field singularity pinpointing them may.

In fact, there is a 't~Hooft-Polyakov-like monopole configuration
in euclidean pure gauge theory, which is an excellent candidate.
It is an extended object which in an appropriate abelian gauge
displays an abelian magnetic point singularity at its centre.
I will show that, by going to another abelian gauge, this gauge field
singularity may be shifted to a different point within the finite extent
of the monopole core.
This provides a nice interpretation of the gauge dependence of monopole
singularities in the abelian projection.

This semiclassical 't~Hooft-Polyakov-like
monopole is the key ingredient in a realization of
the monopole condensation mechanism of confinement in SU(2) gauge theory
proposed some years ago\@.\cite{smitvds}
I will summarize the main results of this work, in particular
an effective action for a class of monopole configurations,
a quantitative estimate of the string tension, and a qualitative
understanding of {\em de\/}confinement.
This will be followed by a brief discussion, in the light of this model,
of numerical results on the effective action and the monopole density.

I would like to emphasize that this realization of the dual superconductor
hypothesis is perfectly compatible with the abelian projection approach.

\section{The dual superconductor hypothesis}

The dual superconductor hypothesis\cite{Ho,Ma} assumes that the
ground state of QCD is a condensate
of chromomagnetic monopoles, causing the chromoelectric flux between
quarks to be squeezed into tubes,
in a similar way as the condensation of electric charges is responsible
for the squeezing of magnetic flux between (imaginary) magnetic charges
in a superconductor.
Such a flux tube behaviour would imply a linear confining potential
between quarks.
In short: condensation of magnetic monopoles causes confinement.

What do we mean by ``monopole condensation''?
I would like to advocate a fairly general definition: monopole condensation
means that monopole configurations dominate, or determine, the ground state,
in the sense that in their absence the ground state would be a different one.
Accordingly, I think one should avoid pushing
the analogy with the superconductor any further than necessary.
For example, the dual superconductor hypothesis would
not \apriori\ require, as sometimes claimed,\
$(i)$ that QCD is described by a dual BCS theory;\
$(ii)$ that monopoles (and antimonopoles) condense in pairs;\
$(iii)$ that the monopole mass is negative;\
$(iv)$ that monopole configurations are local minima of the
action (although it is obviously helpful if they are approximate
stationary solutions, with a large portion of
phase space associated to them);\
$(v)$ that the QCD vacuum is in some sense ``of type II''\@.
(In QCD one does not consider the response to external chromoelectric
fields.
Furthermore, how would the magnetic flux between a magnetic
monopole-antimonopole pair in a type I superconductor be distributed?).

In Ref.~4, for example (cf.~Sec.~6),
the dual superconductor hypothesis
is thought to be realized by certain semiclassical monopole configurations
which dominate due to a large entropy factor which compensates for
the (positive) energy of the monopoles.

\section{The abelian projection and its monopoles}
\label{abproj}

In his influential 1981 paper,\cite{Ho81}\ 't~Hooft outlined a strategy
to implement the dual superconductor idea.
He proposed that in order to investigate a theory a useful first step
would be to identify the physically relevant degrees of freedom.
In a gauge theory, this involves removing the gauge redundancy
without introducing ghosts.
At certain points the gauge fixing condition is ill-defined,
and these gauge-fixing singularities were argued to be unavoidable
and to signal additional, physically relevant dynamical variables.
In the abelian Higgs model, for example, singularities in gauge fixing
to the unitary gauge occur at the location of magnetic vortices, whose
physical significance is beyond dispute.

In a non-abelian gauge theory, it is convenient to leave the largest
abelian subgroup unfixed, as this guarantees the singularities to be
pointlike (linelike in four dimensions);
the remaining abelian gauge freedom can easily be dealt with afterwards.
This is called the ``abelian projection''.\cite{Ho81}
Among these ``abelian gauges'',
the diagonalization gauges, in which a given operator in the adjoint
representation is diagonalized, do not require ghost fields.
't~Hooft also wrote down an abelian gauge which is both Lorentz
covariant and gauge covariant with respect to the abelian subgroup
but does require the introduction of ghost fields.
This gauge is, in schematic notation, given by
\be
(\partial_\mu + i A_\mu^{\rm ab}) \, A_\mu^{\rm non-ab} = 0 .
\label{mag1}
\ee

The singularities in these abelian gauges correspond to
magnetic monopoles with respect to the unfixed abelian subgroup.
{\em However, this does not mean that
these monopoles are singular objects!\/}
The singularity in the gauge field, which enables us to
detect the monopole, is just an artefact of the gauge condition;
the monopole itself can be assumed to be a regular extended object.
The same is true for the vortex in the abelian Higgs model:
it can be characterized by a point (or line) singularity in the field
values, but nonetheless it has a thickness determined by the Higgs mass.

Neither is there a reason to assume that the magnetic charge is pointlike.
In the case of the 't~Hooft-Polyakov monopole, for example,
the monopole core is a region of (approximate) symmetry restoration
and the definition of the unbroken U(1) component becomes ambiguous there.
Only far outside the core is the abelian magnetic field uniquely defined.
This will be illustrated in Sec.~6 by showing that the singularity
can be relocated within the monopole core by going from one abelian
gauge to another.

\section{Abelian dominance\ \ ---\ \ an interpretation}

The lattice implementation of the abelian projection was pioneered
by Kronfeld \etal,\cite{kron1,kron2}\ who studied monopole densities
in SU(2) gauge theory in several diagonalization gauges and in the
so-called maximally abelian gauge (MAG) in which a certain functional
of the gauge fields is maximized\@.\footnote{Imposing the MAG
is equivalent to finding the minimum
of the covariant kinetic action $\int (D_\mu V)^2$ for an $S^2$-valued
spin field $V$, transforming in the adjoint representation,
in the background of the given gauge field
configuration.\protect\cite{thesis}}\ \,
The condition for a stationary point of this functional
reduces to the covariant abelian gauge \rf{mag1}
in the continuum limit.\cite{smitvds}

A very interesting
development was the discovery of ``abelian dominance'' in SU(2)
lattice gauge theory by Suzuki and Yotsuyanagi.\cite{suzabdom}
They observed numerically
that the ``abelian Wilson loop'' in the MAG, that is
the Wilson loop formed by the abelian component of the gauge field
after gauge fixing to the MAG, exhibited area law decay,
$C\exp(-\sigma A)$,
with the same string tension $\sigma$ as the standard (full) SU(2)
Wilson loop.
This kind of behaviour was not observed for the diagonalization gauges.
More recently, it was found\cite{suzmondom,stackmondom}
that the correct string tension is also obtained when the Wilson
loop is calculated from Dirac monopole gauge fields
associated to the monopole currents in MAG-fixed configurations
(``monopole dominance'').
The fact that the MAG appears to identify the abelian, or monopole variables
as the relevant physical degrees of freedom at long distances,
contrary to the diagonalization gauges, is surprising
because the MAG is accompanied by ghosts while the other gauges are not.

What is the meaning of abelian dominance?
A possible interpretation\cite{smitvds} is suggested by the assumption
that the vacuum is dominated by semiclassical 't~Hooft-Polyakov-like
monopoles (cf.\ Sec.~5 and~6).
The abelian monopole field is long-ranged and influences the (full)
Wilson loop at all distances.
The short-ranged non-abelian components, however, can
only interact with the loop when the monopole is close to its perimeter.
The abelian field alone would therefore
cause area law behaviour of the Wilson loop,
whereas the non-abelian fields would only affect the prefactor $C$.

This interpretation will also explain\cite{smitvds}
why a Wilson loop in the adjoint (or in fact any odd-dimensional)
representation of SU(2) does not show area law behaviour.
The abelian gauge field component multiplies $T^3 =$ diag$(1,0,-1)$
in this case, and the Wilson loop reduces to a sum over abelian
Wilson loops with charges $1,0,-1$ (compared with charge $\pm 1/2$ for
the fundamental Wilson loop).
Under the plausible assumption that the charge $\pm 1$ abelian loops
decay with an area law, the constant $\exp(0)=1$ prevents
area law behaviour for the adjoint Wilson loop.
It even excludes perimeter law decay, suggesting that
the perimeter term in the full Wilson loop is due to the
interaction with the non-abelian field components.

\section{A magnetic monopole and its ambiguous singularities}
\label{hpsec}

The euclidean pure SU(2) analogue of the 't~Hooft-Polyakov monopole\cite
{Ho74,Pol74} in the BPS limit is given by (see, \eg, Ref.~4)
\be
A^a_k \ =\ \ep_{aki} \hat x_i \frac{1-K(\mu r)} r,
\ \ \ \ \ \ \ \ \ \
A^a_4 \ =\ \mu \hat x_a \frac{H(\mu r)}{\mu r},
\label{hp1}
\ee
in the ``radial'' or ``hedgehog'' gauge (also satisfying
$\partial_\mu A^a_\mu = 0$),
where $\mu$ is an arbitrary mass scale.
It is a static self-dual configuration, a dyon, which is gauge equivalent
to a periodic instanton\cite{rossi} and carries a topological charge
of $\mu/2\pi$ per unit of (imaginary) time.
The mass of this classical solution is $4\pi\mu/g^2$; see
Ref.~14 for a lattice study of this monopole and its
mass in the quantum theory.
The function $H(\mu r)/\mu r$ goes to 1 for $r\ra\infty$,
and $K$ decays exponentially.
Inspection of the fields for large $r$ shows that this configuration
carries a magnetic charge $-4\pi/g$ with respect to the U(1) subgroup
in the radial direction of SU(2).

The gauge transformation
\be
\Omega(\th,\varphi) \ =\ \exp(-i\varphi T^3)
\, \exp(i\th T^2) \, \exp(i\varphi T^3)
\label{omega}
\ee
rotates this configuration \rf{hp1} into a gauge where $A_4$ is diagonal,
\be
A^{1,2}_k \ =\ f^{1,2}_k (\th,\varphi) \frac K r ,
\ \ \ \ \ \ \ \
A^3_k \ =\ \frac{-\sin \th}{r(1+\cos \th)} \hat\varphi_k ,
\ \ \ \ \ \ \ \
A^a_4 \ =\ \delta_{a3} \frac H r.
\label{hp2}
\ee
Its magnetic nature, now with respect to the 3-direction in group space,
is exhibited more clearly in this gauge:
$A^3_k$ is of the Dirac monopole form
and the non-abelian components
$A^{1,2}_k$ are negligible outside the monopole core.

An important observation\cite{smitvds} is that Eq.~\rf{hp2}
satisfies the abelian gauge condition \rf{mag1}.
The Dirac monopole singularity of $A^3_k$ at the origin is therefore of
the type arising in the abelian projection.
As argued in Sec.~3,
this singularity is just an artefact of the partial gauge fixing condition,
and does not imply that the monopole is a singular object or should be
considered as a point charge;
the presence of
the non-abelian components, which do not vanish for $r\aleq\mu^{-1}$
and are also singular in this gauge, reminds us of the fact that
the definition of the U(1) magnetic field of the monopole makes sense
only outside the monopole core of size $\mu^{-1}$.
Rephrasing this, the Dirac string accompanying the abelian magnetic field
is sharply defined far away, but becomes more and more
``blurred'' as one follows it deeper into the monopole core.
Nevertheless, the selected gauge fixing condition assigns an end point
to the Dirac string: the monopole singularity.
It is this point that is looked for by
algorithms designed to detect magnetic monopoles in lattice
gauge field configurations in abelian gauges, by
monitoring the abelian flux through the faces of elementary lattice cubes.

In view of these considerations, I would like to argue that {\em the
singularity may be found anywhere inside the core of the monopole}.
I will illustrate this with the following example, which may serve
to clarify the statement\cite{Ho81} that not the precise
location of the singularity has a physical meaning, but just its
mere existence.

Consider a static monopole of scale $\mu$ on the lattice.
In the abelian gauge \rf{hp2} the singularity would obviously be
localized at the origin.
Now take the gauge in which $F_{12}$ is required to be diagonal.
Far from the monopole core ($\mu r \gg 1$)
this condition is met by Eq.~\rf{hp2}, with the Dirac string along the
negative $\hat z$-direction,
but in the centre of the monopole ($\mu r \ll 1$) it is satisfied
by Eq.~\rf{hp1}.\cite{stackmondom}
It is then easily checked that the abelian magnetic field at the origin
is weak and that, for
$\mu a$ not too large, the elementary lattice cube at the origin has no
Dirac string coming in, hence ``contains no monopole'' according to the
standard lattice criterion.
To see what happened to the Dirac string and the monopole, note
that the gauge transformation needed
to rotate the monopole from the radial gauge \rf{hp1} into the $F_{12}$
gauge must interpolate between the identity for small $\mu r$ and
$\Omega(\th,\varphi)$ \rf{omega} for large $\mu r$.
Since the singular behaviour of the latter along the negative
$\hat z$-direction is responsible for the Dirac string in the abelian field
(cf.~Eqs.~(\ref{hp1}--\ref{hp2})),
it follows that the end point of the string in the $F_{12}$ gauge
will be assigned to some point on the negative $\hat z$-axis, away
from the origin but inside the core of the monopole.

In conclusion, the (semi)classical extended monopole discussed here is an
excellent candidate for realizing the dual superconductor hypothesis;
it is compatible with the abelian projection and in fact allows one to
understand some of the apparent ambiguities in the definition of
monopole singularities for different abelian gauges.

\section{A realization of dual superconductivity in SU(2) gauge theory}
\label{modelsec}

In Ref.~4 we considered the contribution of a class
of magnetic monopole configurations, of the 't~Hooft-Polyakov form
discussed
in Sec.~5, to the path integral of pure SU(2) gauge theory in
euclidean space.
Condensation of these monopoles was shown to lead to a string tension,
and assuming that the string tension $\sigma$ is dominated by the
contribution of these monopoles, a variational estimate yielded the
value $\sqrt\sigma = 2.3\, \Lambda_{\rm \overline{MS}}
= 45\, \Lambda_{\rm latt}$, in agreement with the number
obtained in lattice Monte Carlo simulations.

This involved the derivation of an effective action for
monopole configurations of arbitrary scale $\mu$, in which the
monopoles were separated by some minimal distance $b \ageq \mu^{-1}$.
This effective action can be written as
\be
S_{\rm eff} \ =\  m_0 b \sum_x k_\mu(x) k_\mu(x)
 + \frac12 \left(\frac{4\pi}{g_R(b)}\right)^2
 \sum_{x,y} k_\mu(x) D(x-y) k_\mu(y),
\label{Seff}
\ee
and
\be
Z_{\rm mon} = \sum_{\{ k_\mu(x)\}}\mbox{}^{\!\!\!\!\prime}\ \exp(-S_{\rm eff})
\label{Zmon}
\ee
is a summation over all configurations of closed monopole loops
on a ``macroscopic'' lattice of spacing $b$.
Here $D(x-y)$ is the propagator, $g_R$ is a running coupling,
and the bare mass term parameter $m_0$
is defined such that $S_{\rm eff}$ reproduces the mass of a static
semiclassical monopole of scale $\mu$.
For more details, see Ref.~4.

This effective action can be mapped on the monopole action of
periodic QED.
The crucial difference with the U(1) monopole action,
besides the bare mass term,
is the presence of a {\em running\/} coupling at the scale $b$,
instead of a bare coupling.
In the U(1) case, monopole condensation occurs for couplings larger
than a critical coupling;
in the present case, it takes place when the
{\em running\/} coupling is large enough, which because of asymptotic
freedom corresponds to large enough $b$.
In the variational estimate of the string tension, where $b$ is taken of the
order of the monopole size $\mu^{-1}$,
this amounts to saying that large enough (or
light enough) monopoles condense and cause confinement.
Along with the earlier-mentioned value of 2.3 for the fundamental ratio
$\sqrt\sigma / \Lambda_{\rm \overline{MS}}$, this estimate produced
the value $\bar\mu = 1.2\,\sqrt\sigma$ for the ``optimal'' monopole scale
$\bar\mu$, with $\bar\mu b = 1.10$,
and a corresponding monopole mass $M = 2.4\,\bar\mu = 2.9\,\sqrt\sigma$.
These ``typical values'', which may be concretized by substituting the
``real-life'' number $\sqrt\sigma\approx 400\,{\rm MeV} =
(0.5\,{\rm fm})^{-1}$,
will be used in the following for a comparison with the monopole density
and the deconfinement transition.

Lattice simulations by Shiba and Suzuki\cite{suzseff} have provided
support for the effective action \rf{Seff}.
These authors computed $S_{\rm eff}$ directly from monopole configurations
derived from Monte Carlo generated SU(2) configurations in the MAG.
The effective action they find has the same form as Eq.~\rf{Seff},
including the bare mass term and a running coupling at a scale $b$.
However, the macroscopic parameter $b$ in their effective action
is defined through a blocking procedure, and the precise relation with
the ``macroscopic'' lattice spacing $b$ in Eq.~\rf{Seff} is not
completely clear.

\section{Monopoles and deconfinement}

This picture of a monopole dominated ground state also led to a qualitative
understanding\cite{smitvds} of deconfinement in terms of monopoles:
{\em At high temperatures the monopoles run predominantly in the
periodic euclidean time direction. This implies deconfinement, with a
transition temperature $T_c \aleq \bar\mu$ for a typical monopole
scale $\bar\mu$.}
The argument goes as follows.
Consider monopoles of some scale $\mu$.
When the temperature $T$ is increased, the length $T^{-1}$
of the euclidean time direction shrinks, and at some point
it becomes smaller than $\mu^{-1}$.
This does not cause any obstruction to timelike monopole configurations,
winding around the short euclidean time direction.
Monopoles whose world line runs in one of the spatial directions, however,
have a spatial extent $\mu^{-1}$ in the euclidean time direction,
such that they cease to ``fit'' in the thin space-time slab.
An additional suppression,
for monopoles with $\mu^{-1} \ageq T^{-1}$,
is presumably caused by a decrease in entropy
relative to the four-dimensional factor $7^L$.
In this connection it is interesting to note
that the deconfinement temperature is of the order of the
typical monopole scale $\bar\mu$;
for $\bar\mu = 1.2\,\sqrt\sigma$ and
$T_c = 0.7\,\sqrt\sigma$ one finds $T_c^{-1}/\bar\mu^{-1} = 1.7$.

The relation with deconfinement is as follows.
Monopoles running in the $y$ and $z$ directions interact with the
Wilson loop in the $x$-$t$ plane and are assumed to be responsible for
its area law decay in the confining phase.
This suggests that such monopoles must be ``suppressed'' in the deconfined
phase, and similarly for monopoles in the $x$-direction.
No such suppression is required for timelike monopoles, which
interact only with Wilson loops in spatial planes.
In fact, spatial Wilson loops show area law behaviour even in the
deconfined phase, and if timelike monopoles are responsible for this,
the $T\ra\infty$ limit
will presumably provide the connection with the three-dimensional picture
of confinement through monopole condensation.\cite{Pol77}

\section{The monopole density}

It is interesting to interpret the results of a lattice
computation\cite{bornetal} of the monopole density in SU(2) gauge theory
in the maximally abelian gauge in the light of Ref.~4.
The aim of this simulation was to investigate whether the monopole density
shows scaling behaviour (independence of the lattice cutoff $1/a$
for sufficiently small $a$),
and whether it has a well-defined infinite-volume limit.
Both properties are important if the monopole density is to be a
physically relevant quantity.

Fig.~1 shows a schematic plot of the monopole
density $\rho$ as a function of $L=Na$,
where $N$ is the number of lattice points in each direction and
$V=L^4$ is the size of the (symmetric) physical four-volume.
In the small-$L$ (small-$a$) region,
data points obtained for different values of
the lattice spacing $a$ and varying $N$
were found to lie on a common curve,
which is evidence for scaling.
For large enough volume, $\rho$ seemed to become independent of the
volume size $L$, although this could not be established rigorously
because these data points were obtained at relatively large values of $a$.
The computed value for $\rho$ corresponds to an average distance of
$(3.3\,\Lambda_{\rm\overline{MS}})^{-1}$ between the monopoles,
which in terms of monopoles of the 't~Hooft-Polyakov type with
$\bar\mu = 1.2\,\sqrt\sg = 2.8\,
\Lambda_{\rm\overline{MS}}$ has the nice interpretation\cite{smitvds}
that the space-time volume is densely packed with monopoles.
\begin{figure}[bth]
\centerline{
\epsfxsize=0.5\textwidth
\epsfbox{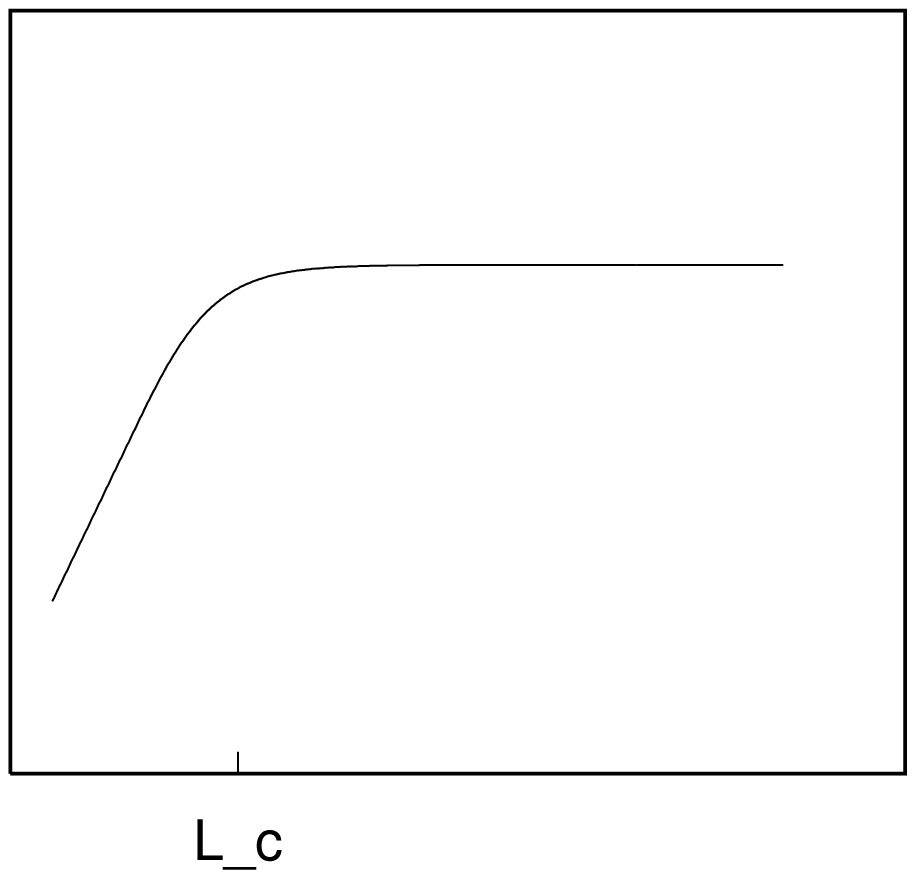}
}
\fcaption{Schematic plot of the monopole density $\rho$ as a function of $L$.
Adapted from Ref.~17.
}
\label{emu3}
\end{figure}

Another interesting observation from Fig.~1 is that $\rho$ increases with
$L$ in the small-volume region before saturating at $L\approx L_c$.
In terms of monopoles of
scale $\bar\mu$
this may be interpreted as follows.\cite{bornetal,suzentr}
For small $L$ ($\bar\mu L \aleq 1$), no monopole or just
one\footnote{We ignore the fact that the number of monopoles
must be even when one adopts periodic boundary conditions.}\ \,
fits into the volume, and due to its energy it is suppressed
compared with the perturbative vacuum.
For intermediate $L$, the entropy factor
is not yet sufficiently large to compensate the energy term.
Only when $L$ becomes of the order of several units of $\bar\mu^{-1}$ can
monopole loops become sufficiently long to have an
exponential entropy factor and does infinite-volume behaviour develop.
This is in agreement with the standard view that asymptotically free
theories in small volumes can be treated perturbatively.

The critical length $L_c$ was found to coincide with the value
at which the Polyakov loop exhibits a jump in symmetric volumes.\cite{kov}
In view of the monopole picture of deconfinement described in Sec.~7,
it is tempting to interpret this as the analogue of the deconfining
transition in a symmetric four-volume, with critical ``box temperature''
$T_c = L_c^{-1}$.
In fact, we observed\cite{bornetal} a peak in the monopole susceptibility
(\ie, the fluctuation in the monopole density) at $L_c$.
However, standard lore does not permit one to speak of phase transitions
in finite volumes, although it may be noted that the continuum limit
$a\ra 0$ at finite physical extent $L=Na$ corresponds to a thermodynamic
limit $N\ra\infty$ in the statistical mechanics sense.

\vspace*{0.3cm}

This work is supported by EC contract ERBCHBICT941067.

\end{document}